\documentstyle[preprint,aps]{revtex}

\newcommand{\be}{\begin{equation}}
\newcommand{\ee}{\end{equation}}
\def\bq{\begin{eqnarray}}
\def\eq{\end{eqnarray}}
\def\n{\nonumber}

\def\mx{\mu_{\xi}}
\def\mxs{\mu_{\xi'}}
\def\mxss{\mu_{\xi''}}
\def\r{\rangle}
\def\l{\langle}

\def\th{\theta}
\def\vp{\varphi}
\def\da{\dagger}
\def\la{\lambda}

\def\lra{\longrightarrow}
\def\ni{\noindent}
\def\hk{{\cal H}_{kin}}
\def\hp{{\cal H}_{phy}}
\def\wt{\widetilde}
\def\wh{\widehat}
\def\hkin{{\cal H}_{kin}}
\def\hphy{{\cal H}_{phy}}

\begin{document}
\draft
\preprint{IMSc/2001/08/48}
\title{Semi-classical States in the Context of Constrained Systems}
\author{G. Date \footnote{e-mail: shyam@imsc.ernet.in}}
\address{The Institute of Mathematical Sciences,
CIT Campus, Chennai-600 113, INDIA.}
\author{Parampreet Singh \footnote{e-mail: param@iucaa.ernet.in}}
\address{Inter-University Centre for Astronomy and Astrophysics,\\
Post Bag 4, Ganeshkhind, Pune-411 007, INDIA.}
\maketitle
\begin{abstract} 
\ni Algebraic quantization scheme has been proposed as an extension of the
Dirac quantization scheme for constrained systems. Semi-classical states
for constrained systems is also an independent and important issue, 
particularly in the context of quantum geometry. In this work we explore this 
issue within the framework of algebraic quantization scheme by means of simple
explicit examples. We obtain semi-classical states as suitable coherent
states a la Perelomov. Remarks on possible generalizations are also
included.
\end{abstract}

\vskip 0.50cm


\narrowtext

\section{Introduction}
Dirac's procedure of quantizing a classical theory with first class
constraints consists of several steps. Firstly one quantizes the system
ignoring the constraints to get a {\it kinematical} Hilbert space, 
$\hkin$. The constraints, represented as self-adjoint operators on $\hkin$
are then imposed as operator equations and physical states are defined
to be the kernel of the constraint operators. The implicit assumption
that physical states belong to the $\hkin$ turns out to be wrong in many
cases of interests and hence a refinement is proposed via the so called
(refined) Algebraic Quantization Scheme 
\cite{alq,abhay,marolf1,marolf2,gm,giulini}. Essentially this includes a
`rigging' of $\hkin$, $\Omega \subset \hkin \subset \Omega^{*}$ and
physical states are sought in $\Omega^{*}$. This allows physical states
to be `distributional' and also allows new physical inner product to be
chosen to define $\hphy$ and physical observables. A map $\eta \ : \
\Omega \ \longrightarrow \ \Omega^{*}$ plays a central role. An example
of such a map is provided by the so called `group averaging procedure'.  \\

There is an independent issue of semi-classical states for a quantum
system. The canonical example of `harmonic oscillator coherent states' 
(standard coherent states), eigenstates of the annihilation operators, 
embodies the idea of semi-classical states. These states are labeled by 
points in the classical phase space $\Gamma$. Furthermore there are observables
(positions and momenta) with respect to which these states are `peaked'
at points in the phase space. This is particularly easy when the phase
space is $R^{2N}$, since the generalized eigenvalues of the positions
and momenta operators provide global coordinates for the phase space.
This is clearly not possible when the phase space is topologically
non-trivial. Such  a phase space can typically be obtained as reduced phase
spaces, $\hat{\Gamma}$ - Constrained surface modulo orbits of the
constraints, and one needs a suitable generalization of the notion of
semi-classical states. \\

Clearly, the first property one needs is that the semi-classical states (in
the quantum Hilbert space) be labeled by points of a classical phase space,
i.e. $|\omega \r,  \omega \in \Gamma$. Second property needed is that of
`peaking'. Given any quantum observable $\hat{F}$, one can immediately
get a function on the classical phase space, $f(\omega) \equiv
\l \omega|\hat{F}|\omega \r / \l \omega|\omega \r$. The idea of peaking is that
we can find enough observables $\hat{F}_i$ such that specifying
$f_i(\omega) = C_i$ will enable one to obtain a unique point,
$\omega(C_i)$, in the phase space. Of course there will be fluctuations:
$\Delta f^2(\omega) = [f^2](\omega) - ([f][(\omega))^2$. These are to be
`small' in a suitable sense eg. `minimum' or within specified windows
$\pm |\delta C_i|$. Clearly one must have at least 2N such observables.
If we can find such $|\omega \r$ and $\hat{F}_i$, then we say that
$|\omega \r$ are candidate semi-classical states. \\

Notice that the notion of semi-classical states itself does
{\it{not}} require any approximation or limiting procedure ($\hbar \to
0$, large quantum numbers etc). These are just states corresponding to
classical states, thus incorporating correspondence principle. 
In principle there could be two or more distinct sets of
semi-classical states. These could be labeled by same phase space or
different phase spaces. The latter case may be construed as an example of
potentially equivalent quantum theories corresponding to two different classical
systems. A requirement that a quantum theory admits 
such semi-classical states is a non-trivial requirement as an arbitrarily 
constructed Hilbert space may or may not admit $|\omega \r, \hat{F}_i$ for 
any choice of a classical phase space. Whether such a notion of semi-classical
states is too permissive or too restrictive is not clear at present. \\

For constrained systems, in practice, it is often convenient to follow
the Dirac quantization procedure (as opposed to the reduced phase space
quantization). The notion of semi-classical states should now be properly
defined in the $\hphy$ and with respect to {\it{physical}} observables.
One may not have as much much control over $\hphy$ as over $\hkin$ as is
the case at present with quantum geometry. One could try to define
semi-classical states in $\hphy$ by first defining them in $\hkin$ and
performing group averaging on them. The peaking property however still
needs to be specified in $\hphy$ using physical observables. Alternatively 
one should obtain a relation between peaking defined relative to physical 
quantities and relative to kinematical quantities. \\

We explore such a strategy in the context of simple toy models with a
single constraint. For the class of models for which physical
observables contain a Lie algebra, one can use corresponding generalized
coherent states a la Perelomov\cite{perel} as candidate semi-classical states. 
Furthermore expectation values of physical observables in $\hphy$ can be 
computed in $\hkin$. \\

The paper is organized as follows: \\

Section II gives a schematic (formal) derivation of the main result. \\

Section III discusses explicit examples implementing the schematic
derivation. The examples are with $\Gamma = R^4$ and a single quadratic
constraint. This has three cases involving compact and non-compact
semi-simple groups. \\ 

Section IV contains remarks on further examples and generalization. A 
discussion of results, possible extensions is also included. 

\section{General Scheme}

Let $\phi$ denote a single constraint (a self adjoint operator on
$\hkin$) and let $G$ be a group commuting with the constraint. Let
$|\xi, k \r$ denote group coherent states labeled by $\xi$ and
constructed from an irreducible representation of $G$ labeled by $k$.
$\xi$ typically denotes points in a coset space while $k$ can be a multi-index 
in general. Clearly $\hkin$ carries a representation, in general reducible, 
of the group and $\phi$ is a multiple of identity on each of the irreducible
blocks. Clearly, the constraint will have a well defined value on every 
irreducible block. Specific value of the constraint will thus select particular
irreducible representation (and possibly copies thereof) labeled by , say,
$\wt k $.\\

In $\hkin$ we have a resolution of identity in the form,

\be
\int d k \, \int d \mx \, |\xi, k\r \l \xi, k| = 1 .
\ee

The integration over $k$ (which can be a sum if $k$ takes discrete
values) is over those values which occur in representations of $G$ in $\hk$
and $d\mu_{\xi}$ is a group invariant measure on a coset space.\\

Following the algebraic quantization scheme, let $\Omega$ be a suitable
dense subspace of $\hkin$ so that we obtain a rigging: $\Omega \subset
\hkin \subset \Omega^* $. For every $|\psi\r \in \Omega$ we have,

\be
|\psi\r = \int d k \, \int d \mx \, \l \xi, k|\psi\r \, |\xi, k\r .
\ee

A map $\eta : \Omega \to \Omega^*$ is proposed to be provided by group
averaging so that 

\be
( \psi| = \frac{1}{V} \int d \la \, \int d k \, \int d \mx \, \l 
\psi|\xi, k\r \l\xi, k| \, e^{- i \, \la \, \hat{\phi}} 
\ee

where $V$ is the group volume, suitably regulated if necessary. We 
denote elements of $\Omega^*$ generically by $( \cdot |~~$ (round bra
instead of angular bra). Now, 

\be
\frac{1}{V} \int d \la \, \l\xi, k| \, e^{- i \, \la \, \hat{\phi}} = 
\delta(k - \wt k) ( \xi, \wt k | .
\ee

Hence we get,

\be
( \psi| = \int d \mx \, \l \psi| \xi, \wt k \r  \, ( \xi, \wt k| .
\ee

The physical inner product, denoted as $\l , \r$ , is defined as 
\cite{abhay,marolf1,marolf2,gm,giulini}

\be
\l \eta \, \psi', \eta \, \psi \r_{phy} = ~~( \psi|\psi'\r .
\ee

The inner product evaluates to

\bq
( \psi| \psi'\r &=& \n \int d k' \, \int d \mxs \, \int d \mx \, 
\l \psi| \xi, \wt k \r  \, ( \xi, \wt k| \xi', k'\r \, \l \xi', k'| \psi'\r \\
&=& \int d \mxs \, \int d \mx \, \l \psi| \xi, \wt k \r  \, \l \xi, \wt k| 
\xi', \wt k\r \, \l \xi', \wt k| \psi'\r .
\eq 

In the first line, we have used resolution of identity on $| \psi' \r$.
Then we use equation (4) and the fact that the constraint operator is 
`block diagonal' with respect to the resolution of identity, to get to
the next line involving only the inner product in $\hkin$. \\

Similarly, the expectation value of a physical observable $\wh A$  is defined 
as

\be
\l \eta \, \psi', \wh A \, \eta \, \psi\r = 
\l \wh A \, \eta \, \psi', \eta \, \psi\r = 
\l \eta \, \wh A \, \psi', \eta \, \psi\r = 
( \psi| \wh A \, \psi' \r
\ee

which evaluates to 

\bq
( \psi| \wh A \psi' \r &=& \n \int d \mx \, \int d \mxs 
\l \psi| \xi, \wt k \r  \, \l \xi, \wt k | \xi', \wt k \r \, 
\l \xi', \wt k | \wh A \, \psi' \r \\
&=& \n \int d k \, \int d \mx \, \int d \mxs 
\, \int d \mxss \l \psi| \xi, \wt k \r  \, \l \xi, \wt k |\xi', \wt k \r \, 
\l \xi', \wt k |\wh A |\, \xi'', k \r \ \l \xi'', k| \psi'\r  .
\eq

Note that for $|\psi \r \in \Omega$, the resolution of identity involves
various representations and thus some of the integrals over the coherent
states labels survive. If however, the kinematical states are chosen as 
$|\psi \r = | \xi_o, \wt k\r$, $|\psi' \r = | \xi^{'}_{o}, \wt k\r$, then 
these integrals can be done. For these choices, the inner product becomes 

\be
\l \eta \, \psi' \, , \eta \, \psi \r = \int d \mxs \, \int  d \mx \, \l \xi_o, \wt k
| \xi, \wt k\r \, \l \xi, \wt k| \xi', \wt k\r \, \l \xi', \wt k
| \xi^{'}_{o} , \wt k \r ,
\ee

which on  using the resolution of identity within an irreducible
representation becomes

\be
\l \eta \, \psi' \, , \eta \, \psi \r = \l \xi_o, \wt k |  \xi^{'}_{o} , \wt k \r = \l \psi| \psi' \r \, .
\ee

This is the statement that if kinematical states are chosen as the coherent 
states of the Lie group generated by a subset of physical observables with the 
representation index selected by the constraint, then the physical inner 
product for the corresponding states is same as the kinematical inner product.\\

The matrix elements of physical observables for the same choice of
states also simplifies in a similar manner and becomes, 

\bq
( \psi| \wh A \psi' \r &=& \n \int d k \, \int d \mx \, \int d \mxs 
\, \int d \mxss \l \psi| \xi, \wt k \r  \, \l \xi, \wt k |\xi', \wt k \r \, 
\l \xi', \wt k |\wh A |\, \xi'', k \r \ \l \xi'', k| \psi'\r  \\
&=&\n \int d \mx \, \int d \mxs 
\l \xi_0, \wt k| \xi, \wt k \r  \, \l \xi, \wt k |\wh A| \xi', \wt k \r \, 
\l \xi', \wt k | \xi'_0, \wt k \r \\
&=& \l \xi_o, \wt k| \wh A| \xi^{'}_{o}, \wt k \r \, = \, \l \psi| \wh A| 
\psi'\r \, .
\eq

Evidently, the fluctuations in physical observables in $\hk$ and $\hp$ are 
related as
\be
 ( \Delta \wh A^2 )_{phy} = ( \Delta \wh A^2 )_{kin} .
\ee

Eqs. (10-12) are the key results. The observation is that {\it 
if one uses the coherent states of a selected representation of the Lie group 
generated by a subset of the physical observables, then the inner product, 
expectation values of and the quantum fluctuations in physical
observables, computed with reference
to these are identical whether computed in the $\hp$ or $\hk$.} Several 
remarks are in order. \\

{\underline {Remarks:}} \\

(1) There is no mention of semi-classical states in the above. The result
is strictly a property of coherent states. Even here properties really used
are the resolution of identity, labeling of coherent states by some
coset space, coherent states being constructed per irreducible
representation. In particular, peaking property is not referred to.
Group averaging is also used only to the extent that it selects a
particular irreducible representation. Although group averaging behaves
as though a projection operator, it does {\it not} give a state in
$\hkin$ in general. In particular we do {\it not} assume that group average of
a coherent state belongs to the $\hkin$. In the final expressions we did 
assume that a selected representation is contained in $\Omega$. Constraint 
is of course used to admit a group $G$ whose coherent states have been used. \\

Semi-classical states can now be introduced with the further assumption namely
{\it the reduced phase space can be (set theoretically) mapped into the coset 
space labeling the coherent states.} The peaking properties of coherent states 
will then give the peaking properties of the semi-classical states. Incidently,
if in addition, the Hamiltonian of the system is the constraint itself, as is 
the case for canonical gravity in the cosmological context, then preservation 
of the peaking properties under time evolution is automatic. \\

It could be that for a particular choice of $G$, one may not get the
desired correspondence. As long as there exist {\underline a} choice of
$G$ commuting with the constraint and {\underline a} choice of
representation selected by the constraint which is labeled by the
reduced phase space, we do have a class of semi-classical states. \\

Note that understanding the full $\hphy$ and all the physical observables is
also {\it not} essential for the identification of semi-classical states. \\

(2) Potential problems with group averaging when the constraint group is
non-compact, can be bypassed as long as any regularization procedure
adopted preserves the representation selection property (eqn. 4). As such one
could translate the implications of group averaging as a condition on
the rigging map and on the choice of $\Omega$. \\

(3) Observe that the coherent states which give the simpler result
depend on the group $G$ which depends on the constraint. One does {\it
not} start with a fixed set of coherent states in $\hkin$ and define the
physical ones via an explicit group averaging. In the examples discussed
in the next section, the contrast will become apparent. \\

\section{Examples}
In this section we consider some simple toy models to illustrate the schematics 
discussed above. The chosen constraints are quadratic in phase space coordinates
and momenta. Each of the case has some distinct feature. We will identify the 
group (of canonical transformations) commuting with the constraint, the
reduced phase space and show the correspondence between the reduced phase space 
and the coherent space labels. In all the cases considered, the
classical phase space is $\Gamma = R^4$ and we {\it choose} $\hkin$ to
be the usual Hilbert space of square integrable functions on $R^2$. For
a more general and detailed analysis of the first two examples, please
see \cite{alq}.

\subsection{The two dimensional harmonic oscillator constraint}
The constraint can be written down as\footnote{In the following the index on coordinates is subscripted only for notational convenience.}
\be
\phi = \frac{1}{2}(q_1^2 + p_1^2 + q_2^2 + p_2^2 - R^2)
\ee

where $R^2$ is positive. On using 

\be
q_i = \sqrt{\frac{\hbar}{2}} (a_i + a_i^\da), \hspace{0.5cm} p_i = -\, i \,  \sqrt{\frac{\hbar}{2}} (a_i - a_i^\da)
\ee

the constraint can be rewritten as

\be
\phi = \hbar \, \left( a_1^\da \, a_1 + a_2^\da \, a_2 + 1 - \frac{R^2}{2 \hbar}\right).
\ee

Among the physical observables permitted by this constraint are:

\bq
J_x &=& \frac{1}{2} \, \left(q_1 \, q_2 + p_1 \, p_2 \right) = \frac{\hbar}{2} \, (a_1^\da \, a_2 + a_2^\da \, a_1), \\
J_y &=& \frac{1}{2} \, \left( q_1 \, p_2 - q_2 \, p_1 \right) = \frac{i \, \hbar}{2} \, (a_2^\da \, a_1 - a_1^\da \, a_2), \\
J_z &=& \frac{1}{4} \, \left(q_1^2 + p_1^2 - q_2^2 - p_2^2 \right) = \frac{\hbar}{2} \, (a_1^\da \, a_1 - a_2^\da \, a_2)
\eq 

which as can be easily seen to form the SU(2) Lie algebra,

\be
[J_x, J_y] = i \, \hbar \, J_z , \hspace{0.5cm}  
[J_y, J_z] = i \, \hbar \, J_x , \hspace{0.5cm}  
[J_z, J_x] = i \, \hbar \, J_y .
\ee 

The Casimir invariant $(C_2)$ for the SU(2) group is $J^2 = J_x^2 + J_y^2 + 
J_z^2$ with eigenvalues $\hbar^2 j(j + 1)$, which in our case becomes,

\be
C_2 = \frac{\hbar^2}{4} \, \left[ \frac{1}{\hbar^2} \left( \phi + 
\frac{R^2}{2} \right)^2 - 1 \right] = \hbar^2 j(j + 1) .
\ee

When the constraint is imposed, it becomes, 

\be
C_2 =  \frac{1}{4} \, \left[\frac{R^4}{4 \hbar^2} - 1\right] = j(j + 1) .
\ee

The solution of $j$ for the above equation is the representation of the physical
coherent state which is selected by the group averaging procedure described in 
previous section. \\

Our next step is to identify the correspondence between the points on the 
reduced phase space and the points which serve as labels for the SU(2) 
coherent states. For that we rewrite the constraint as

\be
\frac{q_1^2 + p_1^2}{R^2} + \frac{q_2^2 + p_2^2}{R^2} = 1
\ee

which suggests a convenient parameterization,

\bq
q_1 = R \, \cos\th \, \cos \vp_1 , &\hspace{1cm}& p_1 = R \, \cos\th \, \sin \vp_1 \\
q_2 = R \, \sin\th \, \cos \vp_2 , &\hspace{1cm}& p_2 = R \, \sin\th \, \sin \vp_2 .
\eq

The SU(2) group elements are parameterized as:

\be
\pmatrix{\alpha & \beta \cr  -\beta^* & \alpha^*\cr}, 
\hspace{0.3cm} |\alpha|^2 + |\beta|^2 = 1, 
\hspace{0.2cm} \alpha \equiv \cos \frac{\mu}{2} \, e^{- i \, \nu_1}, \, 
\beta \equiv -\, \sin\frac{\mu}{2} \, e^{- i \, \nu_2} 
\ee

where $0 \leq \mu \leq  2 \pi$ and $0 \leq \nu_1, \nu_2 \leq 2 \, \pi$. Hence,
the mapping between the constrained surface and the group manifold is given by 
$\mu/2 \lra \th, \, - \nu_1 \lra \vp_1, \, - \nu_2 \lra \vp_2$. \\

The physical observables are related to the parameters on the constrained 
surface as 

\be
\frac{J_y}{J_x} = \tan (\vp_2 - \vp_1), \hspace{0.5cm} J_z = \frac{R^2}{4} ( 2\,  \cos^2\th - 1)
\ee 

Introducing $\vp_{\pm} \equiv (\vp_1 \pm \vp_2)/2 $, one sees that 
trajectories of the constraint just change $\vp_+$. The reduced phase space 
is thus parameterized by $\theta, \vp_-$. Putting $\vp_+ = 0$, we obtain the 
mapping between the reduced phase space and the coset space as $\mu/2 \lra \th, 
- \nu  \lra \vp$, where we have redefined $\nu := \nu_2$ and $\vp := \vp_-$. \\

The SU(2) coherent states are labeled by $\zeta$ given in terms of the 
coset space labels as

\be
\zeta = -\, \tan\frac{\mu}{2} \, e^{-i \, \nu}
\ee 

and thus we get a mapping between the classical reduced phase space and 
the coherent states labels. \\

The SU(2) coherent states are the minimum uncertainty states of any pair of 
$J_x$, $J_y$ and $J_z$,

\be
|\zeta, j \r =  \sum_{m = -j}^{j} \, \left[\frac{(2 \, j)!}{(j + m)! \, (j - m)!}\right]^{1/2} \, (1 + |\zeta|^2)^{-j} \, \zeta^{j + m} \, |j,m\r
\ee

with the resolution of identity,

\be
\int d\mu (\zeta, j) |\zeta, j\r \, \l \zeta, j| = 1, \hspace{1cm} d\mu (\zeta, j) = \frac{2 j + 1}{\pi \, (1 + |\zeta|^2)^2} \, d^2 \zeta .
\ee

Since all the requirements of our general scheme are satisfied group averaging 
picks out the representation corresponding to $j = (- 1 + R^2/2 \hbar)/2$. 
The states of this representations are the ones which give the peaking with
respect to  the physical observables. \\

{\underline {Remarks:}} \\

(1) The reducible representation of SU(2) in $\hkin$ contains all allowed
values of $j$. This can be inferred from the known spectrum of the
two dimensional harmonic oscillator Hamiltonian. The fact that the $j$ can 
only take discrete values, implies that in this case the group averaging 
produces physical states in $\hkin$ itself. It also implies that if $R$ does 
not have a value so as to select an allowed $j$, then there are {\it no} 
physical states either in $\hkin$ or in a $\Omega^*$. Thus $R$ itself must 
take only a set of allowed values. For an analysis of this example from
a different view point see also \cite{radhika}.\\

(2) In this case one could have used the standard coherent states of
the oscillator and constructed physical states corresponding to these by
direct explicit group averaging. The physical state comes out to be

\be
|z_1, z_2\r = \int \, e^{- i \lambda \, (R^2/2 \hbar - 1)} \, |z_1 \, e^{i \lambda}, z_2 \, e^{i \lambda}\r \, d \lambda
\ee

where 

\be
|z\r = e^{-|z|^2/2} \, \sum_{n = 0}^{\infty} \, \frac{z^n}{\sqrt{n !}} \, |n\r
\ee

and $z$ is the eigenvalue of the annihilation operator 
($a =  (q + i \, p)/\sqrt{2 \hbar}$). \\

In this case it turns out that $\l J_i \r_{phy}$ are equal to their respective 
classical expressions, which is not surprising because the constraint under 
consideration is the harmonic oscillator Hamiltonian operator. \\

(3) The mapping between the points on the reduced phase space and the points 
which label the coherent states is important in our analysis. That establishes  
the correspondence between classical and quantum regimes. We will see that this 
is possible in all the cases considered. 

\subsection{The out of phase harmonic  oscillator constraint}

The constraint is of the form 

\bq
\phi &=& \n \frac{1}{2}(q_1^2 + p_1^2 - q_2^2 - p_2^2 - R^2) \\
&=&   \hbar \, \left( a_1^\da \, a_1 - a_2^\da \, a_2  - \frac{R^2}{2 \hbar}\right).
\eq

The physical observables forming a Lie algebra in this case are

\bq
K_x &=& \frac{1}{2} \, \left(q_1 \, p_2 + q_2 \, p_1  \right) = \frac{i \, \hbar}{2} \, (a_1^\da \, a_2^\da  -  a_1 \, a_2),\\
K_y &=& \frac{1}{2} \, \left( q_1 \, q_2 -  p_1 \, p_2 \right) = \frac{\hbar}{2} \, (a_1^\da \, a_2^\da + a_1 \, a_2), \\
K_z &=& \frac{1}{4} \, \left(q_1^2 + p_1^2 + q_2^2 + p_2^2 \right) =\frac{\hbar}{2} \, (a_1^\da \, a_1 + a_2^\da \, a_2 + 1)
\eq 

which  form the SU(1,1) Lie algebra,

\be
[K_x, K_y] = -i \, \hbar \, K_z , \hspace{0.5cm}  [K_y, K_z] = i \, \hbar \, K_x , \hspace{0.5cm}  [K_z, K_x] = i \, \hbar \, K_y .
\ee 

The Casimir invariant for SU(1,1) group is: 

\be
C_2 = K_z^2 - K_x^2 - K_y^2 .
\ee

Its eigenvalues are $\hbar^2 k(k - 1)$, for discrete 
series and $\hbar^2 (-\lambda^2 - \frac{1}{4}) $,  for 
continuous series. For the discrete series $k > 0$. For the continuous series 
to have a representation such that coherent states are labeled by a 
coset space one must have, $\lambda > 0$ for the principal continuous series 
and $- i \sigma/2 < \lambda < i \sigma/2, \lambda \ne 0$ for the supplementary 
continuous series. The Casimir invariant for this constraint turns out to be

\be
C_2 = \frac{\hbar^2}{4} \left[(N_1 - N_2)^2 - 1\right] = \frac{\hbar^2}{4} \, \left[ \frac{1}{\hbar^2} \left( \phi + \frac{R^2}{2} \right)^2 - 1 \right] 
\ee

where $N_i = a^\da _i \, a_i$. Then, it is easy to see that the only allowed 
series in this case is the discrete series and all members of this
series occur in the reducible representation of SU(1,1) in $\hkin$.  
Upon imposition of the constraint one gets,

\be
C_2 =  \frac{1}{4} \, \left[\frac{R^4}{4 \hbar^2} - 1\right] = k(k - 1) \hspace{0.3cm} .
\ee

The parameters on the constraint surface can be identified by rewriting the 
constraint as,

\be
\frac{q_1^2 + p_1^2}{R^2} - \frac{q_2^2 + p_2^2}{R^2} = 1
\ee

which suggests, 

\bq
q_1 = R \, \cosh\xi \, \cos \vp_1 , &\hspace{1cm}& p_1 = R \, \cosh\xi \, \sin \vp_1 \\
q_2 = R \, \sinh\xi \, \cos \vp_2 , &\hspace{1cm}& p_2 = R \, \sinh\xi \, \sin \vp_2 .
\eq

Further, the SU(1,1) group elements are parameterized as,

\be
\pmatrix{\alpha & \beta \cr  \beta^* & \alpha^*\cr}, \hspace{0.3cm} 
|\alpha|^2 - |\beta|^2 = 1, \hspace{0.3cm}  
\alpha \equiv \cosh \frac{\tau}{2} \, e^{- i \, \nu_1}, \, 
\beta \equiv \sinh \frac{\tau}{2} \, e^{- i \, \nu_2} .
\ee

where $\tau > 0$ and $0 \leq \nu_1, \nu_2 \leq 2 \, \pi$. Hence, the mapping 
between the parameters on the constraint surface and those on group manifold is
$\tau/2 \lra \xi, \, - \nu_1 \lra \vp_1, \, - \nu_2 \lra \vp_2$. \\

The physical observables are related to the parameters of the constraint 
surface as,

\be
K_z = \frac{R^2}{4} \, (2 \, \cosh^2 \xi - 1), \hspace{1cm} \frac{K_x}{K_y} = \tan (\vp_1 + \vp_2).
\ee

Introducing $\vp_{\pm} \equiv (\vp_1 \pm \vp_2)/2$, one sees that the
trajectories of the constraint involve only changes in $\vp_-$. The
reduced phase space is thus parameterized by $\xi, \vp_+$. Putting $\vp_-
= 0$, one obtains a mapping between the reduced phase space and the coset 
space (in this case it is the Lobachevsky plane) labeling the coherent
states as, $\tau/2 \lra \xi,  - \nu \lra \vp$, where we have redefined 
$\nu := \nu_2$. \\

The SU(1,1) coherent space are labeled by points on the coset space  as

\be
\zeta = \tanh \frac{\tau}{2} \, e^{-i \, \nu} ~~~~~,~~~~~ |\zeta| < 1 .
\ee

With the above identifications we  obtain the correspondence between the 
classical reduced phase space and the coherent states. \\

The SU(1,1) coherent states are also the minimum uncertainty states for 
the physical observables $K_x$ and $K_y$, for the discrete series they are

\be
|\zeta, k \r = (1 - |\zeta|^2)^k \, \sum_{n = 0}^{\infty} \, \left[\frac{\Gamma(n + 2 k)}{\Gamma(n + 1) \, \Gamma(2 k)} \right]^{1/2} \, \zeta^n \, |k, n\r
\ee

with the following resolution of identity,

\be
\int d \mu(\zeta, k) \, |\zeta ,k\r \, \l \zeta, k| = 1, \hspace{1cm} d \mu(\zeta, k) = \frac{2 k - 1}{\pi \, (1 - |\zeta|^2)^2} \, d^2 \zeta .
\ee
 
Group averaging picks out the representation\footnote{Another representation is picked out if $ R^2 < 2 \, \hbar . $} 
with $k = (1 + R^2/2 \hbar)/2$. \\

{\underline {Remarks:}}\\

(1) Since in this case coherent states correspond to a discrete series, the
comments made in the context of SU(2) apply here also. Thus quantization is 
possible only if $R$ is discrete. \\

(2) One could have used the standard coherent states in this case too. The 
physical state comes out to be

\be
|z_1, z_2\r = \int \, e^{- i \lambda \, R^2/2 \hbar } \, |z_1 \, e^{i \lambda}, z_2 \, e^{-i \lambda}\r \, d \lambda.
\ee

However, in this case one recovers the classical values for only $K_x$ and 
$K_y$. \\

(3) In the case of SU(1,1) coherent states the resolution 
of identity is not defined for $k < 1/2$, however, this problem can be averted 
by  resorting to the weak resolution of identity \cite{brif}. 

\subsection{The two dimensional inverted oscillator}

The constraint in this case is

\bq
\phi &=& \n \frac{1}{2}(-\, q_1^2 + p_1^2 - q_2^2 + p_2^2 - R^2) \\
&=&   -\frac{\hbar}{2}  \, \left( a_1^{\da 2} + a_2^{\da 2} + a_1^2 + a_2^2 
 + \frac{R^2}{ \hbar}\right).
\eq

The physical observables corresponding to the above constraint  are

\bq
K_x &=& \frac{1}{4} \, \left(q_1^2 - p_1^2 - q_2^2  + p_2^2 \right) = 
\frac{\hbar}{4} \, (a_1^{\da 2} - a_2^{\da 2}  +  a_1^2  - a_2^2),\\
K_y &=& \frac{1}{2} \, \left( q_1 \, q_2 -  p_1 \, p_2 \right) = 
\frac{\hbar}{2} \, (a_1^\da \, a_2^\da + a_1 \, a_2) \\
K_z &=& \frac{1}{2} \, \left(q_1 \, p_2  - q_2 \, p_1 \right) = 
\frac{i \, \hbar}{2} \, (a_2^\da \, a_1 - a_1^\da \, a_2 )
\eq 

which  also form an SU(1,1) algebra. The Casimir invariant for this case is

\be
C_2 =    -\frac{\hbar ^2}{4} \, \left[ \frac{1}{ \,\hbar^2} 
\left( \phi + \frac{R^2}{2} \right)^2 + 1 \right]. 
\ee

The representations allowed in this case are corresponding to the principal
continuous series. Upon imposition of the constraint, the Casimir becomes

\be
C_2 =  -\frac{1}{4} \, \left[\frac{R^4}{4 \hbar^2} + 1\right] =  
-(\lambda^2 + \frac{1}{4}) .
\ee

To get the identification with SU(1,1) group manifold, we rewrite the 
constraint as,

\be
\frac{ p_1^2 + p_2^2}{R^2} - \frac{q_1^2 + q_2^2}{R^2} = 1
\ee

which suggests a parameterization, 

\bq
p_1 = R \, \cosh\mu \, \cos \gamma_1 , &\hspace{1cm}& 
p_2 = R \, \cosh\mu \, \sin \gamma_1 \\
q_1 = R \, \sinh\mu \, \sin \gamma_2 , &\hspace{1cm}& 
q_2 = R \, \sinh\mu \, \cos  \gamma_2 .
\eq

Hence, the identification between the group manifold and coset labels for 
SU(1,1) coherent states is $\tau/2 \lra \mu$ and $-\nu \lra 
(\gamma_1 + \gamma_2)$. \\

Note that on the constraint surface $(- \, q_1^2 \, + \, p_1^2)$ and $(q_2^2 
\, - \, p_2^2)$ are constant. Furthermore, we can rewrite $K_x$ and 
$K_z^2 - K_y^2$ as sum and product of these,

\be
K_x = -\frac{1}{4} \, \left[ (-\, q_1^2 + p_1^2) + (q_2^2 - p_2^2) \right], 
\hspace{1cm} 
K_z^2 - K_y^2 = \frac{1}{4} \, \left[(-q_1^2 + p_1^2) \,(q_2^2 - p_2^2) \right].    
\ee

In order to get the mapping  between the reduced phase space and the coset 
space  we  write the constraint in the form which will give natural parameters
for the constraint surface,

\be
\frac{(- \, q_1^2 + p_1^2)}{R^2} - \frac{(q_2^2 - p_2^2)}{R^2} = 1
\ee

which leads to three possibilities: \\

(i) $(- \, q_1^2 + p_1^2) > 0$ and $(q_2^2 - p_2^2) > 0$ , 
\par (ii) $(- \, q_1^2 + p_1^2) < 0$  and  $(q_2^2 - p_2^2) < 0$ , 
\par (iii) $(- \, q_1^2 + p_1^2) > 0$ but $(q_2^2 - p_2^2) < 0$ . \\

\par The fourth possibility of $(- \, q_1^2 + p_1^2) < 0$  and 
$(q_2^2 - p_2^2) > 0$ is ruled out because $R^2$ is always positive. \\

{\underline{Remark:}}\\

 The three cases are mutually exclusive. Each of these 
constitutes a connected component of the reduced phase space. One can restrict 
one self to any one of these. In the following, all three are considered
but it is to be remembered that only one is relevant at a time. \\

For each of these cases there is a natural parameterization in terms of $(\xi, \eta_1, \eta_2)$, 

case (i):

\bq
q_1 = R \, \cosh\xi \, \sinh \eta_1 , &\hspace{1cm}& 
p_1 = R \, \cosh\xi \, \cosh \eta_1 \\
q_2 = R \, \sinh\xi \, \cosh \eta_2 , &\hspace{1cm}& 
p_2 = R \, \sinh\xi \, \sinh \eta_2 ,
\eq

case (ii):

\bq
q_1 = R \, \sinh\xi \, \cosh \eta_1 , &\hspace{1cm}& 
p_1 = R \, \sinh\xi \, \sinh \eta_1 \\
q_2 = R \, \cosh\xi \, \sinh \eta_2 , &\hspace{1cm}& 
p_2 = R \, \cosh\xi \, \cosh \eta_2 ,
\eq

and case (iii):
\bq
q_1 &=& R \, \mbox{sech} \, {\it \xi} \, \sinh \eta_1 , \hspace{1.1cm} 
p_1 = R \, \mbox{sech} \, {\it \xi} \, \cosh \eta_1 \\
q_2 &=& R \, \tanh\xi \, \sinh \eta_2 , \hspace{1cm} 
p_2 = R \, \tanh\xi \, \cosh \eta_2 .
\eq

In case (iii), note that the choice of the use of $\sin \, \th$ and $\cos \, 
\th$ instead of $\mbox{sech} \, \xi$ and $\tanh \xi$ is not suitable because 
then we would have to omit points  $\th = (0, \pi)$ and/or $(\pi/2, 3 \pi/2)$. 
The reduced phase space is hence made up of disconnected parts. \\  

The physical observables can be expressed in terms of $(\xi, \eta_1, \eta_2)$ 
and one finds out that the parameters on the reduced phase space are $\xi$ and 
$(\eta_1 - \eta_2)$. For example, in the case (i) we get,

\be
K_x = \frac{R^2}{4} \, (1 - 2 \,  \cosh^2 \xi), \hspace{1cm} 
\frac{K_y}{K_z} = \tanh (\eta_2 - \eta_1).
\ee

With the parameterization of the group manifold, coset space and reduced phase 
space given, it can be shown that there is a 1-1 and onto mapping between the 
coset space and the reduced phase space. We will show this for case (i), other 
cases follow similarly. \\

We first define $\eta_{\pm} = (\eta_1 \pm \eta_2)/2$. Further, note that 
$\dot \eta_+ = 1$ and $\dot \eta_- = 0$. Without any loss of generality we can 
choose $\eta_+ = 0$ to get to the reduced phase space. With these substitutions 
in eqs(60,61) and then equating the resultant set with eqs.(56,57), we get

\bq
q_1 &=& R \, \cosh \xi \, \sinh\eta = R \, \sinh \mu \, \sin \gamma_2 \\  
p_1 &=& R \, \cosh \xi \, \cosh \eta = R \, \cosh \mu \, \cos \gamma_1 \\
q_2 &=& R \, \sinh \eta \, \cosh \eta = R \sinh \mu \, \cos \gamma_2 \\
p_2 &=& - R \, \sinh \xi \, \sinh \eta = R \cosh \mu \, \sin \gamma_1
\eq

where $\eta = \eta_-$. The above set of equations leads to,

\be
\sinh 2\mu \, \sin (\gamma_1 + \gamma_2) = \sinh 2\eta, \hspace{0.2cm} 
\sinh 2\mu \, \cos(\gamma_1 + \gamma_2) = \sinh 2\xi \, \cosh 2\eta. 
\ee

With our new definitions $(\xi, \eta)$ parameterize the reduced phase space and
as noted earlier $\mu$ and $(\gamma_1 + \gamma_2)$ are related to the coset 
space labels for the SU(1,1) coherent states $(\tau/2 \lra \mu$ and $-\nu \lra 
(\gamma_1 + \gamma_2))$. Hence, the above equation gives us a mapping between 
the reduced phase space and the coset space. That the  mapping is 1-1 and onto 
is easy to check, since given either of  $(\mu, (\gamma_1 + \gamma_2))$ or 
$(\xi, \eta)$, one can determine other uniquely. This can be shown similarly 
for other two cases. \\

Hence, reduced phase space is made up of the three copies of the same space 
which is equivalent to the coset space. Which copy is to be picked out is
to be {\it predetermined} classically as remarked above. \\

Since, in this case coherent states corresponding to principal continuous 
series are allowed, they are given by expansions in orthonormal basis with 
expansion functions as the eigenfunctions of Laplace-Beltrami operator for 
the Lobachevsky plane. We refer the reader to Perelomov's monograph\cite{perel}
for more details. The representation which is picked out by group averaging 
is $\lambda = R^2/4 \hbar$. \\

{\underline {Remark:}}\\

As in the previous example, we have the same non-compact group here,
but now the representations in $\hkin$ belong to the principal continuous
series. The use of the standard coherent states leads to messy algebra in obtaining
the group averaged states, obscuring any correspondence between quantum and 
classical states. However, the natural choice of coherent states of the 
invariance group simplifies the computations considerably.

\section{Discussion}
We have looked at the simplest (smallest dimensional) non-trivial 
possibilities for a constrained system. Our examples involve both compact 
and non-compact 
semi-simple groups. In all cases, the symmetry group was the group of
linear canonical transformations leaving the constrained surface
invariant. The constrained surface itself was a group manifold. Dimensions
of the reduced phase spaces and the coset spaces labeling the coherent
states also coincided. This need not always be the case. For instance if
one considers the first example generalized to $N$ dimensions, one
encounters SU(N) as the symmetry group. The constrained surface is
$S^{2N -1}$ which is not a group manifold for $N > 2$. If one uses the SU(N) 
coherent states given in \cite{nemoto}, then these are labeled by (2N -1)
dimensional coset space ($\frac{SU(N)}{SU(N -1)} \sim S^{2N -1}$) while the 
reduced phase space is of dimension (2N -2). Semi-classical states then form a 
{\it proper subset} of the coherent states. The schematics of section II
of course applies to the coherent states. This is a case in which the
reduced phase space is mapped {\it in to} the coset space labeling
the coherent states. \\ 

Another example which can be commented upon is the case with $\Gamma =
R^{2N}$ and constraint being $q_N = 0$ (say). The choice of $G$ in this
case would be the Heisenberg-Weyl group in the $(2N -2)$ phase space
variables. The reduced phase space is trivially $R^{2N -2}$ and the
standard coherent states will suffice. The physical inner product among
these is automatically restricted to the $(N -1)$ dimensional Lebesgue
measure. \\

While there are very many interesting examples that can be analyzed, eg.
the case of a free relativistic particle constraint, 
$P_0^2 - \vec{P}\cdot\vec{P} = m^2$ and its non-relativistic cousins,
it is not our aim to do so here.  We have been primarily motivated by the
quantum geometry context wherein the issue of semi-classical states is
being addressed. One issue in this regards is the need to identify
semi-classical states in $\hphy$ on the one hand and having good control
only over $\hkin$ on the other hand. Our schematic derivation addresses
this issue though with further assumptions made. \\

The toy models considered in this work bring out various points one should 
keep in mind while obtaining physical semi-classical states through refined 
algebraic quantization. \\

(1) The use of coherent states for the invariance group allows us to establish 
the result that the inner product, expectation values of and quantum 
fluctuations in physical observables are same for the kinematical and physical 
states, if kinematical states are chosen as the group coherent states with a 
specific representation. We have explicitly demonstrated this in the three 
non trivial examples and it is apparent that this can be done whenever we have 
an invariance group  and suitable coherent states available.  \\

(2) The mapping between the reduced phase space and the coset space labels 
which define the coherent states is generally non trivial. This has been 
seen in the case of an  inverted oscillator. This is a point  to be kept in 
mind because apart from the simple problems in most cases one is not likely 
to know the reduced phase space sufficiently explicitly. \\

(3) While the availability of group averaging as a method of choosing 
$\eta$ mapping is tempting to suggest that one could start with a notion of 
semi-classical states in $\hkin$ and group average these to {\it define} 
semi-classical states in $\hphy$, it can be extremely cumbersome. In our 
examples this corresponds to choosing the standard coherent states as starting 
states and implementing group averaging. We have seen how clumsy this is.
Therefore an important lesson is that one should not over-emphasize the use of 
standard coherent states (or coherent states natural to the structure of
$\hkin$) in all situations. While the standard coherent states are a natural 
choice for harmonic oscillator constraint they prove quite unsuitable for
the inverted oscillator constraint. \\

We should emphasize that we have {\it not} taken the examples 
as illustration of the algebraic quantization scheme. We are also
{\it not} using coherent states to quantize a constrained theory i.e. construct 
the full physical state space. We are using coherent states with a limited 
purpose as a means to identify candidate {\it semi-classical} states. Towards 
this end, particular choice of $\hkin$ and precise details of group averaging 
are not too critical. \\

Our scheme also has some limitations. It relies on the availability of a
suitable Lie group (as opposed to only a Lie algebra) commuting with the 
constraint. Even if a group is available, it relies on availability of suitable 
coherent states. Lastly it relies on finding or demonstrating existence of a 
suitable map between the label space for the coherent states and the physical 
reduced phase space. Typically coherent states are labeled by a (connected) 
coset space i.e. a connected homogeneous space whereas a reduced phase space 
need not be so. It is however a conceivable possibility that reduced phase 
space may map in to only a subset of coherent states. Each of these aspects 
needs to be explored and generalized further to get to interesting realistic
systems. \\

However, if these limitations can be circumvented then one has a 
possibility of getting the semi-classical states at least without having to 
know the full physical space. \\

{\underline{Acknowledgements:}} PS is supported by CSIR grant number: 2 - 34/98(ii)E.U-II. He thanks the Institute of Mathematical Sciences for warm hospitality and a pleasant stay during which this work was completed.

\end{document}